\documentclass[11pt]{article}
\usepackage{latexsym}
\usepackage{amssymb}

\def\o{\omega}
\def\bra{\langle}
\def\ket{\rangle}

\def\d{\delta}
\def\D{\Delta}
\def\e{\epsilon}

\def\f{\phi}
\def\k{\kappa}
\def\l{\lambda}
\def\L{\Lambda}

\def\half{\frac{1}{2}}
\def\m{\mu}
\def\n{\nu}
\def\s{\sigma}
\def\t{\tau}
\def\o{\omega}
\def\r{\rho}

\def\th{\theta}

\def\vf{\varphi}

\def\dd{\mbox{d}}
\newcommand{\ti}[1]{\tilde{#1}}

\newcommand{\sm}[1]{\mbox{\scriptsize #1}}

\renewcommand{\@}[1]{\sqrt{#1}}

\def\be{\begin{eqnarray}}
\renewcommand{\le}[1]{\label{#1}\end{eqnarray}}
\def\ee{\end{eqnarray}}
\newcommand{\bea}{\begin{eqnarray}}
\newcommand{\eea}{\end{eqnarray}}
\newcommand{\eq}[1]{(\ref{#1})}
\def\nn{\nonumber\\}

\def\ffract#1#2{\raise .35 em\hbox{$\scriptstyle#1$}\kern-.25em/
\kern-.2em\lower .22 em \hbox{$\scriptstyle#2$}}

\def\pa{\partial}

\def\half{{1\over2}}

\newdimen\tableauside\tableauside=2.0ex
\newdimen\tableaurule\tableaurule=0.8pt
\newdimen\tableaustep
\def\phantomhrule#1{\hbox{\vbox to0pt{\hrule height\tableaurule width#1\vss}}}
\def\phantomvrule#1{\vbox{\hbox to0pt{\vrule width\tableaurule height#1\hss}}}
\def\sqr{\vbox{%
  \phantomhrule\tableaustep
  \hbox{\phantomvrule\tableaustep\kern\tableaustep\phantomvrule\tableaustep}%
  \hbox{\vbox{\phantomhrule\tableauside}\kern-\tableaurule}}}
\def\squares#1{\hbox{\count0=#1\noindent\loop\sqr
  \advance\count0 by-1 \ifnum\count0>0\repeat}}
\def\tableau#1{\vcenter{\offinterlineskip
  \tableaustep=\tableauside\advance\tableaustep by-\tableaurule
  \kern\normallineskip\hbox
    {\kern\normallineskip\vbox
      {\gettableau#1 0 }%
     \kern\normallineskip\kern\tableaurule}%
  \kern\normallineskip\kern\tableaurule}}
\def\gettableau#1 {\ifnum#1=0\let\next=\null\else
  \squares{#1}\let\next=\gettableau\fi\next}

\begin{document}
\begin{flushright}
AEI-2005-172\\
UPR-1138-T\\
\end{flushright}
\vskip0.1truecm

\begin{center}
\vskip 2.5truecm {\Large \textbf{A Diffusion Model for $SU(N)$ QCD Screening}}
\vskip 1truecm

{\large \textbf{Giovanni Arcioni${}^\star${}, Sebastian de Haro${}^\S$, and Peng Gao${}^\ddagger$}}\\

\vskip .7truecm
$^\star${\it The Racah Institute of Physics\\
The Hebrew University\\
91904 Jerusalem, Israel}\\
{\tt arcionig@phys.huji.ac.il}
\vskip 4truemm ${}^\S${\it Max-Planck-Institut f\"{u}r Gravitationsphysik\\
Albert-Einstein-Institut\\
14476 Golm, Germany}\\
\tt{sdh@aei.mpg.de}
\vskip 4truemm ${}^\ddagger${\it Department of Physics and Astronomy\\
University of Pennsylvania\\
Philadelphia, PA 19104-6396, USA}\\
\tt{gaopeng@physics.upenn.edu}
\vskip 4truemm

\end{center}

\vskip 2truecm

\begin{center}

\textbf{\large Abstract}

\end{center}

We consider a phenomenological model for the dynamics of Wilson
loops in pure $SU(N)$ QCD where the expectation value of the loop is the
average over an interacting diffusion process on the group manifold
$SU(N)$. The interaction is provided by an arbitrary potential that
generates the transition from the Casimir scaling regime into the
 screening phase of the four-dimensional gauge theory. The potential is required
to respect the underlying center symmetry of the gauge theory,
and this predicts screening of arbitrary $SU(N)$ representations to the corresponding antisymmetric
representations of the same $N$-ality. The stable strings before the onset of screening are therefore the $k$-strings.
In the process we find a non-trivial but solvable modification of the
QCD$_2$ matrix model that involves an arbitrary potential.

\newpage

\tableofcontents

\section{Introduction}

Our current understanding of the quark confinement mechanism is
based on the assumption of a linearly confining potential with color
electric flux between fundamental charges localized in a tube
connecting them. Wilson loops are the typical gauge invariant
operators, i.e. observables, one considers: they represent the order
parameters which allow to detect the presence of a confining phase.
It is therefore of extreme importance to study their dynamics. Loop
equations were already formulated in an elegant way long ago
\cite{Makeenko:1979pb} but it turns out to be extremely difficult to
solve them exactly, since Wilson loops are quite complicated objects
themselves being non-local and composite (See \cite{Polyakov:2004vp}
for a very interesting review of the topic and connections with
string theories).

Numerical evaluations of Wilson loops are therefore very useful to shed
light on the problem. Indeed, extensive computations in lattice
gauge theory support quark confinement and are able to describe
quite in detail the behavior of the static quark potential. Recent
studies \cite{Shevchenko:2003ks,DelDebbio:1995gc,Bali:2000un,
Lucini:2001nv,deForcrand:1999kr} have shown, in
particular, the presence of an intermediate region where the string
tension of the linearly confining potential is proportional to the
quadratic Casimir operator (of the color group); hence the name
``Casimir scaling". However, as soon as the distance between the
charges increases, at some point screening by gluons becomes
energetically favored and Casimir scaling breaks down
\cite{Lucini:2001nv,Kallio:2000jc}. We just want to recall
that screening in general just shows up as a slight deviation from
Casimir scaling and that one can also consider charges in higher
representations of the color group rather than the fundamental. For
the $SU(2)$ and $SU(3)$ cases, any such charge can be screened
either to the fundamental or to the trivial representation; for the
case of $SU(N)$ with $N>3$, however, there can be -stable- strings
with string tensions different from the fundamental one, which are
normally referred as
$k$-strings.

It would be desirable, of course, to give an exhaustive and more
clear explanation of these results from lattice gauge theory
simulations. It is well-known, in particular, that Casimir scaling is
exact in two dimensional Yang Mills theories for all kind of loop sizes.
In this specific case, however, confinement is a perturbative result
and de facto the distribution of flux is random, since the
plaquettes decouple. In addition there is no transition to the
screening regime. Some years ago, however, Ambjorn et al
\cite{Ambjorn:1984dp,Ambjorn:1984mb} made, the suggestions
that four dimensional QCD, at large distances, reduces to two
dimensional Yang Mills because of a sort of ``dimensional reduction".
The proposal was motivated by the physics of spin systems in
random magnetic fields which displays such a behavior. In QCD then
one assumes that a necessary and sufficient condition for
confinement is a vacuum with random fluxes. The latter play the role
of the magnetic field to give effectively dimensional reduction,
which could then explain Casimir scaling. This proposal, as we will
comment below, was already tested numerically in
\cite{Belova:1983rd,Makeenko:1982gr} and the results of the
numerical simulations show good agreement. Still, however, one has
 to explain the screening of the charges. Ensembles of Wilson loops, in particular, were
 considered and their probability distribution was computed.

A phenomenological model along these lines was proposed in
\cite{blnt}, where it is assumed that the probability distribution
precisely undergoes diffusion on the group manifold of the
corresponding color group. The model is in principle able to explain
the transition from Casimir to screening regimes and the analysis
was carried out for the $SU(2)$ case. In our paper we extend the
analysis to the $SU(N)$ case and by demanding explicitly center
symmetry of our dynamical system  we can see the emergence of
$k$-strings as well. We will translate this proposal in terms of a
matrix model which is a non-trivial perturbation of the 2d Yang-Mills matrix
model where area is no longer preserved under
diffeomorphisms when going from Casimir to the screening
regime. At the same time we give a more robust mathematical
description of the model.

The organization of the paper is as follows: in Section 2 we review
how to relate the Wilson loop distributions to diffusion processes.
In particular we recall the assumption of dimensional reduction. 
In Section 3 we proceed along the lines of \cite{blnt}
revisiting the free diffusion on the group manifold and generalizing
it to the $SU(N)$ case. We show the emergence of Casimir scaling. In
Section 4 we add a drift term discussing briefly some of its
properties and map the problem to an equivalent quantum mechanical
path integral in the presence of a potential. In Section 5 we demand
invariance under center symmetry and see how the model is able to
explain the transition from Casimir scaling to the screening regime.
We also show the emergence of $k$-strings and some specific choices
for the potential. In Section 6 we discuss a candidate effective
action to describe the Wilson loop dynamics. We consider then the
diffusion process at small times and make connections with weakly
coupled Yang Mills along the lines of \cite{blnt}. In Section 7 we
comment on the vortex density energy which can be predicted from the
model and make some links with lattice gauge theory (LGT). We end
with some conclusions and open problems. In the Appendices we
collect some group theory definitions and properties we used
throughout the paper.

\section{Wilson loop distributions as diffusion processes}

We are first going to review some of the ideas underlying the
approach we are following. As said, indeed, according to \cite{blnt}
one associates the Wilson loop distribution to a diffusion process
on a group manifold. Similar proposals, however, already appear in
the literature, as we are now going to revisit
briefly in this section.

Indeed it has already been noticed in
\cite{Ambjorn:1984dp,Ambjorn:1984mb,Belova:1983rd,Makeenko:1982gr}
that the value $w$ of a Wilson loop undergoes
Brownian motion as the size of the loop is progressively
increased. In addition, each such path as a function of the area
represents a single random walk and an ensemble of such trajectories
gives the spectral density $\r$ of the values $w$ of the Wilson loop
 \be
\r(w) = \bra0 \mid\delta(W-w)\mid0\ket
 \ee
Qualitatively, as soon as the size of the Wilson loop becomes larger,
$\r$ becomes uniform: the gauge field is weakly correlated and
the distribution of eigenvalues represents disordered
configurations. Most notably,
confinement is then an outcome of such uniform distribution.

 To carry out such analysis the hypothesis of ``dimensional
reduction" was explicitly assumed and tested. The latter was
originally suggested in \cite{Ambjorn:1984dp},
\cite{Ambjorn:1984mb}. More precisely, dimensional reduction is a
consequence of the assumption of ``stochastic confinement", which
states that the QCD vacuum is made of a disordered color magnetic
fluxes. One can then infer that the IR dynamics of four dimensional
QCD is given by {\it two} dimensional QCD with quite good
approximation.
For another derivation of dimensional reduction see
\cite{Greensite:1979yn}. Dimensional reduction is a consequence, in
any case, of the specific stochastic assumption on the nature of the
QCD {\it vacuum}.  Note also that stochastic confinement implies Casimir
scaling, which was conjectured long ago before recent lattice
simulations.

Dimensional reduction was then tested in
\cite{Ambjorn:1984dp,Ambjorn:1984mb,Belova:1983rd,Makeenko:1982gr}.
Qualitatively, at weak
coupling (up to usual roughening transition problems), dimensional
reduction seems to be reproduced. At strong coupling, however, a
more quantitative and remarkable result was obtained
\be
\rho_{I\times J}^{d=4} (\alpha, \beta) \sim \rho_{I \times J}^{d=2}
(\alpha, \beta_{2d})
\ee
namely the spectral density $\rho$ of a
Wilson loop of size $I \times J$ in four dimensions at coupling
$\beta$ depending on the value of the eigenvalue $\alpha$ (consider
here for simplicity the $SU(2)$ case where there is only one eigenvalue
as discussed in more in detail below) is roughly equal to the {\it
two} dimensional one evaluated at a coupling $\beta_{2d}$ which is a
function of the four dimensional one $\beta$. Note that the two
spectral densities almost coincide already with the renormalization of
one single parameter. Summarizing: one is studying a two dimensional
effective model where the ``residua" of the four dimensional asymptotic
freedom are in the dependence of the two dimensional coupling
$\beta_{2d}$ from the four dimensional one $\beta$. In the approach
of \cite{blnt} this effective model is described by a diffusion
process on the group manifold and the dependence from the coupling
is reabsorbed in the ``time" $t$ entering the
diffusion equation.

The proposal of \cite{blnt}, in particular, is to identify the
spectral density $\r$ with the kernel of the the heat equation on the group manifold of the
color group. Computations have been performed assuming
somehow the dimensional reduction already at work, since they turn
out to be computable in a deformed version of QCD$_2$, where character
expansions on the group manifold are exact. Actually it is also well
known that the partition function of 2d Yang-Mills satisfies a heat
equation on the group manifold so the proposal to identify the
spectral density $\r$ with the heat kernel solution on the
group manifold of the color group is not surprising, {\it once}
dimensional reduction has been invoked. The same conclusion is
obtained from the geometrical lattice gauge theory (LGT) actions
discussed in Section 7 where we believe however it comes out in a
more transparent way. We want to stress, however, that our theory is
{\it not} QCD$_2$ but a perturbation of it, and the non-trivial physics
lies in this perturbation.

\section{Free diffusion of Wilson loops on $SU(N)$ and Casimir scaling}

We are first going to review the free diffusion of the spectral density
of Wilson loops $\r$ for the $SU(2)$ case which was
already discussed in \cite{blnt}. We then generalize to $SU(N)$.
Free diffusion will give Casimir a scaling regime, i.e. the
string tension of the confining linear potential is proportional to
the quadratic Casimir of the color group.

Casimir scaling in general holds, approximately, at intermediate
ranges, from onset of confinement till color screening
\footnote{Roughly till $1.2$fm \cite{Bali:2000un}.}. Theoretical
arguments in favor of it are dimensional reduction as discussed and
also Witten analysis in the large $N$ limit \cite{cargesewitten},
where Casimir scaling becomes exact. Numerical evidence was provided
quite recently by lattice simulations
\cite{DelDebbio:1995gc,Bali:2000un}.

\subsection{Revisiting free diffusion on $SU(2)$}

This is the case discussed in \cite{blnt}. It is useful to review it
and make some additional remarks before moving to the more general $SU(N)$ case.

The probability distribution of Wilson loops is given in this case
by\footnote{In our case,  $G$ is the spectral
density $\r$. Compare with \cite{Makeenko:1982gr}: their
$\r={2\over\pi}G$. In addition our time $t$ will depend both on the
coupling and the size of the loop as a consequence of the model we
are considering. We give some explicit examples of this dependence
in Section 6.} 
\be\label{P} 
P(\theta,t) = \sin^2\theta\, G(\theta,t)
 \ee
where $G(\theta,t)$ is the kernel of the heat equation on
$SU(2)=S^3$, i.e. it satisfies
\be
 \left({\pa\over\pa t}-\Delta\right)G(\th,t)={1\over\sin^2\th}\,\d(\th)\d(t)~.
\ee
In this case $\theta$ is the first polar angle on $S^3$ while $t$ is
time. The Laplacian on $S^3$ is:
\be
\D_{S^3}={1\over2\sin^2\th}\,\pa_\th\left(\sin^2\th\,\pa_\th\right)
+{\sin^2\th\over2\sin\f}\,\pa_\f\left(\sin\f\,\pa_\f\right)
+\half\sin^2\th\sin^2\f\,\pa_\vf^2~,
\ee
where the factor of $\half$ is conventional. There is an $SO(4)$ symmetry group
acting on the $S^3$, and the
eigenfunctions of this Laplacian are the ${\cal Y}_{nlm}(\th,\f,\vf)$,
$l=1,\ldots,n$, $m=-l,\ldots,l$, with eigenvalues
\be
\D_{S^3}{\cal Y}_{nlm}=-\half\, n(n+2)\,{\cal Y}_{nlm}~,
\ee
and $n=0,1,2,\ldots$
The Cartan subalgebra of $SU(2)$ however only depends on a maximal torus,
parametrized by $\th$. Assuming gauge invariance of the background and of
the Wilson loop operators fixes the remaining angles $\f$ and $\vf$ to constant
values. Thus, we will be considering
$\th$-dependent functions only. Instead of using the ${\cal Y}_{nlm}$'s,
it will then be more convenient to use a basis $SU(2)$ characters, which are
eigenfunctions of the ``radial" part of the Laplacian:
\be
\D_{S^3}\chi_j(\th)=-2j(j+1)\chi_j(\th)
\ee
where $n=2j$. The $SU(2)$ character is:
\be
\chi_j(\th)={\sin(2j+1)\th\over\sin\th}~.
\ee



The same will go through for $SU(N)$. Again one can
invoke gauge invariance arguments. This restricts
then the dependence of $G$ only on $N-1$ ``radial" coordinates of the Cartan
subalgebra. We
know that from the point of view of LGT, as we will see in section 7,
this is natural, since any function one builds up out of group
elements has to be a class function in order to preserve gauge
invariance, i.e. it has to depend only on radial coordinates. This
reduction to ``flat space" namely the maximal torus spanned by the
$N-1$ Cartan generators has been already pointed out in
\cite{Dowker1} where quantum mechanics on group manifolds was
discussed. There the reduction to radial coordinates  can be
interpreted geometrically as rotational invariance in group space
sense. In more mathematical language this is related to the so
called intertwining operators \cite{Anderson:1989db}.

Using standard spectral decomposition methods the kernel $G$ is
easily found to be
\be\label{G}
 G(\th,t)={2\over\pi}\,\th(t)\sum_{n=1}^\infty n{\sin n\th\over\sin\th} e^{-(n^2-1)t/2}~.
\ee
We have shifted $n$ by one unit, so $n=2j+1$ and we see that the exponent indeed falls off
with the Casimir. We now compute Wilson loops in this Brownian motion ensemble average.
One easily finds finds
\be\label{casimirscaling}
\frac{\log \bra W_{j_1}\ket}{\log\bra W_{j_2}\ket} = \frac{j_1(j_1+1)}{j_2(j_2+1)}
\ee
where $\bra W_j\ket$ is the vev of the Wilson loop operator in the spin
$j$ representation of $SU(2)$. One sees therefore that by definition
we are in the Casimir scaling regime. Note also that as $t
\rightarrow \infty $ $G \rightarrow 2/\pi$, i.e. one has a uniform
distribution which is an indication of confinement as recalled in
Section 2.

\subsection{Free diffusion on $SU(N)$}

We now generalize the diffusion process to the $SU(N)$ case.
As explained above, gauge invariance allows us to restrict the Laplacian
to the Cartan subalgebra, and we get the following free heat equation:
\be\label{freede}
\left({\partial\over\partial t}-\Delta\right)G(t,\theta)={1\over
J^2(\theta)}\,\delta(t)\d(\theta)~.
\ee
Here $\theta=\theta_i=(\theta_1,...,\theta_N)$ are the Cartan
coordinates, subject to the $SU(N)$ constraint:
\be
\sum_{i=1}^N\th_i=0~;
\ee
in the following we will always assume this constraint and also drop the subscript from
$\theta_i$. $J(\theta)$ is the
Jacobian associated with $SU(N)$ Haar measure. The explicit expressions and more details
of the solution are given in Appendix \ref{kernels}.

Let us consider the more general problem
\be \left({\pa\over\pa
t}-\D\right)g(t;\th,\th')={1\over J^2(\th)}\,\d(t)\d(\th-\th')~. \ee
The right-hand side gives us the choice of boundary condition at
$t=0$. The solution therefore is: \be\label{gexpansion}
g(t;\th,\th')=\th(t)\sum_\l\chi_\l(\th)\chi_\l(\th')\,e^{-\half\,C_2(\l)t}~.
\ee where the sum runs over all irreducible $SU(N)$ representations,
labeled by $\l$. In turn, this is a sum over integers, as in the
simple $SU(2)$ case \eq{G}. Details are explained in Appendix
\ref{group}.

The above is obviously the partition function of 2d Yang-Mills on
the cylinder, where $t=g_{\sm{2dYM}}^2A$ where $A$ is the area.
Sending $\th=\th'\rightarrow0$, we get the partition
function of QCD$_2$ on the sphere. Both theories are well-known to
have phase transitions in the large $N$ limit. In this paper we will
be concerned with the case $\th'=0$. In that case, the above
correlator corresponds to 2d Yang-Mills on the disc: 
\be
g(t,\th)=\th(t)\sum_\l\chi_\l(\th)\,\mbox{dim}(\l)\,e^{-\half\,C_2(\l)t}~.
\ee 
$g(t,\th)$ can also be seen as a Brownian motion probability $p(t,\th)$:
\be\label{sunensemble}
g(\theta,t) =  p(\theta,t) ~. \ee
This relation will get modified in the interacting case.

We will insert a Wilson loop operator with holonomy specified along
the boundary of the disc, so the picture where we identify the
partition function of 2d Yang-Mills with the Casimir dynamics
dimensionally reduced to the surface spanned by the loop is correct
in the free case. In the interacting case to be discussed in the
next sections this picture will essentially remain unchanged; only
the theory on this surface will no longer be pure 2d Yang-Mills.

The Wilson loop average is now given by computing the expectation
value of the Wilson loop operator in the ensemble \eq{sunensemble}:
\be\label{Wloop} 
\bra W_\l(\th)\ket=\int[\dd\th]\,W_\l(\th)\,p(t,\th)~. 
\ee 
Here and in what follows $\dd\th$ will
be understood to be the Haar measure on $SU(N)$:
$[\dd\th]=\dd^{N-1}\th J^2(\th)$. Let us now discuss what the Wilson
loop operator should be. Clearly, gauge invariance requires it to be
a class function of $\th$. That means that we can expand it in a
basis of $SU(N)$ characters. Further, the holonomy around the loop
singles out a particular representation. Therefore, up to an overall
normalization constant we get
\be\label{Wloopop}
W_\l(\th)=\chi_\l(\th)~.
\ee
This
is of course a usual fact in 2d Yang-Mills. However, notice that the above argument
applies to the interacting theory as well. We can now easily compute
its expectation value in the free case: 
\be\label{casimirscaling2}
\bra W_\l(\th)\ket=\mbox{dim}(\l)\,e^{-\half\,C_2(\l)t}~, \ee 
where
we used standard properties of characters, summarized in Appendix
\ref{group}. Notice that this directly leads to Casimir scaling as
in the $SU(2)$ case, \eq{casimirscaling}. The factor of
$\mbox{dim}(\l)$ is an irrelevant normalization constant, which
depends on our overall choice of normalization in \eq{Wloopop}. The
crucial fact here really is that the probability $p(t,\th)$ is
correctly normalized, 
\be \int[\dd\th]\,p(t,\th)=1~, 
\ee 
so that the ensemble \eq{sunensemble} is a good Brownian motion ensemble.

All of the above is completely standard in 2d Yang-Mills. When we include the potential, we will see that
the representation of the holonomy of the Wilson loop will no longer be the only one contributing to its expectation value.
There will be a mixing where all representations of the same $N$-ality will contribute, and this is of course what makes
the dynamics of the interacting theory non-trivial.


\section{Diffusion with a potential and screening}

\subsection{Drift picture}

As we have seen above, free diffusion on $SU(N)$ reproduces Casimir
scaling. In QCD, however, the external charges interact with
intermediate strings that can be in arbitrary representations. This
is reflected by the fact that Wilson loop averages are no longer of
the type \eq{casimirscaling2}. We expect states in arbitrary
representations to contribute to the final answer for the
expectation value. Thus, we need to generalize the free diffusion
equation \eq{freede}. An interaction term \cite{blnt}, namely a small drift
term $V(\th)$, will give us a Fokker-Planck equation that is no longer diagonal:
\be
{\pa\over\pa t}p(t,\th)=\D p(t,\th)+\pa_i\left(\pa^iV\,p(t,\th)\right)~.
\ee
The fact that the interaction is given by a drift term ensures that there is a
conserved current and the heat equation takes the form of a
continuity equation. This is standard and we will not dwell on it.

We explained above that the dynamics of the theory mixes the representations that contribute to a Wilson
loop. The contributing representations are not arbitrary though; there is a restriction coming from
their $N$-ality. $N$-ality is the number of boxes of the Young tableau modulo $N$, see Appendix \ref{group}
for the precise formula. For example, it is well-known that for large loops the leading contribution is that of
lowest energy, which is the antisymmetric representation of fixed $N$-ality. Thus the behavior is quite
different from \eq{casimirscaling2}.
We will see that this $N$-ality dependence follows from a drift that respects the center symmetry of the
background.

Let us  briefly review the $SU(2)$ case discussed in \cite{blnt} before we do the $SU(N)$ generalization.
The diffusion equation with drift term can be rewritten as:
\be\label{diffeqdrift}
\left({\pa\over\pa t}-\ti\D\right)p(t,\th)={1\over\sin^2\th}\,\d(t)\d(\th)
\ee
where we incorporated the boundary condition as a source on the right-hand side, and
\be\label{tiD}
\ti\D=\D+\D V+\pa^iV\pa_i~.
\ee
is the modified Laplacian. Note that this operator is no longer hermitian.

To solve this equation we proceed as before; we expand in a basis of
eigenfunctions $\ti\psi_n$ like in \eq{gexpansion}:
\be\label{conditions}
p(t,\th)=\th(t)\sum_n\ti\psi_n(\th)\ti\psi_n(0)e^{-E_nt+V(0)} \ee
where the constant term was added for convenience. The new solutions
$\ti\psi_n$ are eigenfunctions of the modified Laplacian:
\be\label{eigenval}
\ti\D\ti\psi_n=-E_n\ti\psi_n~, \ee and the
boundary condition requires \be
\sum_n\ti\psi_n(\th)\ti\psi_n(0)e^{V(0)}={1\over\sin^2\th}\,\d(\th)~.
\ee

The eigenvalue equation \eq{eigenval} amounts to
\be\label{lapl}
{1\over\sin^2\th}{\pa\over\pa\th}\left(\sin^2\th{\pa\over\pa\th}\ti\psi_n+\sin^2\th{\pa
V\over\pa\th}\,\ti\psi_n\right) =-E_n\ti\psi_n~.
\ee
It has an obvious zero energy solution:
\be
\ti\psi_0=e^{-V(\th)}~.
\ee
Notice therefore that the asymptotic distribution in the diffusion
process is no longer uniform on $S^3$ as in the free case.

As discussed, we now require that the drift respects center symmetry
(we will explain the role of center symmetry in more detail in section 5).
We show in Appendix \ref{group} that the center acts on $\th$ by
a shift $\th\rightarrow\th+\pi$ when $N=2$. The key fact here is that the Wilson loop
operator transforms under center symmetry according to its $N$-ality:
\be
W_j(\th+\pi)=(-)^{2j}W_j(\th)~.
\ee
Loops in integer and in half-integer representations will thus be screened
differently. In general, however, the drift term does mix the representations.
If we want it not to mix them, it must have the same symmetries
as the background:
\be
V(\theta+\pi)=V(\theta)~.
\ee
This is nothing than reflection symmetry around the equator\footnote{This of course reflects the
fact that $SU(2)$ is the double cover of $SO(3)$. In the usual $SO(3)$ representation
where $\vf$ is the rotation angle around the azymuthal axis of the two-sphere,
$\th=\half\,\vf$. $\vf$ has periodicity $2\pi$, which is why $\th$ only has periodicity
$\pi$.}.

The above symmetry gives at asymptotic times (i.e. strong coupling) screening of the
integer Wilson loops to adjoint loops and screening of half integer loops
to fundamental loops \cite{blnt}.
The role of this symmetry will be simply to keep, at
asymptotic times, states with different $N$-ality distinct.
Note however that the presence of the drift by itself does not
ensure a perimeter law. In the $SU(2)$ case, one
expects a perimeter law for the integer representations (this would
correspond indeed to a $k=0$ string). Still one has area
scaling as can be checked. To implement the perimeter scaling one
should generalize and require an explicit time dependence of the
drift. In that case, asymptotically, one could impose \cite{blnt}:
\be
V(\theta,t) \rightarrow V_{\infty} (\theta) e^{-m\sqrt{t}} \ee with
$m$ mass of the ``gluelumps".

Before moving to the $SU(N)$ case it will be useful to map the diffusion
problem in the presence of drift to a diffusion process in the
presence of a potential, to be thought of in path integral language.
It is  well known that a Markov process with some specified drift
can be mapped to a quantum system in path integral theory. Here we
simply follow the same route working however in the Euclidean at the
level of the quantum system. This mapping will be explicitly used in
the next section. We first review and give a slight generalization of
the $SU(2)$ case discussed in \cite{blnt}.

\subsection{The screening potential}

The transformation that maps the diffusion problem with a drift to the quantum mechanical problem of
a particle in a potential is easy to find:
\be\label{wavefrescaling}
\ti\psi_n(\th)=e^{-\half V(\th)}\psi_n(\th)~.
\ee
It is easy to check that the diffusion equation
\be\label{diff2}
\left({\pa\over\pa t}-\ti\D\right)p(t;\th,\th')={1\over\sin^2\th}\,\d(\th-\th')\d(t)
\ee
becomes
\be
\left({\pa\over\pa t}-\D+U(\th)\right)g(t',t;\th',\th)={1\over\sin^2\th}\,
\d(\th-\th')\d(t-t')
\ee
where the potential is given in terms of the drift as:
\be\label{potentialU}
U(\th)=-\half\,\D V+{1\over4}(V')^2=e^{-\half V(\th)}\D e^{\half V(\th)}~.
\ee
The rescaling of the wave-functions \eq{wavefrescaling}
also induced a rescaling of $p(t,\th)$ according to:
\be\label{p}
p(t,\th)=g(0,t;0,\th)\,e^{-\half\,V(\th)+\half\,V(0)}~.
\ee
Using \eq{wavefrescaling}, we get:
\be
g(t',t;\th',\th)=\th(t-t')\sum_n\psi_n(\th)\psi_n(\th')\,e^{E_n(t'-t)}~.
\ee
This also explains our choice of overall scaling constant in \eq{conditions}.

The above analysis straightforwardly generalizes to the $SU(N)$ case.
The heat equation with drift \eq{diffeqdrift} can again be solved
in terms of new eigenfunctions $\ti\psi_\l$:
\be\label{eqp}
p(t,\th)=\th(t)\sum_\l\ti\psi_\l(\th)\ti\psi_\l(0)e^{-E_\l t+V(0)}~.
\ee
The sum ranges over all irreducible $SU(N)$ representations, which is again
a sum over integers as explained in Appendix \ref{group}.

As before, there is a non-uniform zero mode
\be
\ti\psi_0=e^{-V(\th)}~.
\ee
By the same mapping \eq{p}, we get:
\be\label{diffeq2}
\left({\pa\over\pa t}-\D+U(\th)\right)g(t,t';\th,\phi)
={1\over J^2(\th)}\,\d(\th-\phi)\d(t-t')~,
\ee
with the potential
given by
\be
U(\th)=e^{-\half V(\th)}\D e^{\half V(\th)}~.
\ee
Again, \eq{diffeq2} is solved by
\be\label{solutiong}
g(t,t';\th,\phi)=\th(t-t')\sum_\l\psi_\l(\th)\psi_\l(\phi)\,e^{E_\l(t-t')}~,
\ee
where $\ti\psi_\l$ and $\psi_\l$ are related by the same wave-function
rescaling \eq{wavefrescaling}. The zero mode is
\be
\psi_0(\th)=e^{-\half V(\th)}~.
\ee
Plugging \eq{solutiong} back into \eq{diffeq2}, we get the conditions on the
eigenfunctions:
\be\label{qm}
\left(\D-U(\th)\right)\psi_\l(\th)=-E_\l\psi_\l(\th)~.
\ee
Our task in what follows will be to solve this equation, subject to the boundary
condition that follows from \eq{diffeq2}:
\be\label{conditions}
\sum_\l\psi_\l(\th)\psi_\l(\th')={1\over J^2}\,\d(\th-\th')~.
\ee
In the next section we show that the quantum mechanical model \eq{qm} is
exactly solvable, and explain how the solution is connected to the
physics of string $N$-ality.

\section{Center symmetry the emergence of $k$-strings}

We have seen that the presence of a small drift term and/or
equivalently of a potential can bring about small deviations from
Casimir scaling which are expected when screening of the sources takes
place. We are going to develop this point further and show
explicitly the emergence of the $k$-strings. The only assumption one
has to make is invariance of the potential and/or drift
term under center symmetry.

To that end, let us briefly recall the role of
the $N$-ality of the center of the color group. In our case
the color group is $SU(N)$ and the center is $Z_N$. An element
of $Z_N$ will act on a $SU(N)$ representation via a factor
\be
\exp\left(\frac{2 \pi i n k}{N}\right).
\ee
We call $k$ the $N$-ality of the
representation.  Now it turns out that fields which can give string
breaking are the ones with $N$-ality different from zero. Therefore
confinement can be stated as follows: start from pure gauge $SU(N)$
theory plus matter and consider those fields with $N$-ality
different from zero. Send their masses to
infinity. Confinement amounts to show that in this limit there
exists a confining phase in which the work needed to separate a
quark and an anti-quark to a distance $L$ tends to $\sigma L$ where
now the string tension $\sigma$ depends on the representation. In the screening
phase, in particular, that is to say after string breaking, it will
depend only on the $N$ ality of the
representation. 

In general, matter with non zero $N$-ality breaks the global center
symmetry of the lagrangian of the theory, which is required by the
standard order parameters (Wilson loops, Polyakov loops, 't Hooft
loops and similar) to be able to detect the confined phase. This is
why the assumption of invariance under center symmetry is natural
as well in the case of the diffusion process
which is supposed to describe different regimes in the
confined phase.


\subsection{String tension from the diffusion process}

We now come to the core of the problem, which is the $SU(N)$ generalization
of the interacting diffusion process.

We should first discuss how the center of the gauge group acts on the
Wilson loop operator. In the $SU(N)$ case, this operator is the
$SU(N)$ character in representation $\l$, \eq{Wloopop}. The center acts on it as:
\be\label{center}
W_\l(\th_i+{2\pi\over N})=e^{2\pi ik(\l)/N}W_\l(\th_i)~,
\ee
where $k(\l)$ is the $N$-ality of the representation.
The proof of this is given in the Appendix \ref{group}. The background and the potential,
however, are invariant under center symmetry, as already discussed.
This means that the drift satisfies:
\be\label{sympot}
V(\th_i)=V\left(\th_i+{2\pi\over N}\right)~.
\ee
Hence, since the background and the
potential are invariant under center symmetry,
the above property \eq{Wloopop} is a fundamental property of the loop that
distinguishes  states of different $N$-ality. More precisely,
we find asymptotically:
\be\label{Wasymptotics}
\bra
W_\l(\th_i)\ket\rightarrow C\, e^{-E_\m t}~,
\ee
with some proportionality constant $C$ which will depend on the
representations. $E_\m$ is the tension of the string. For
example, for $SU(2)$, $E_\m=0$ for $\l=j\in\mathbb{Z}$, whereas it
is non-zero if $j\in \half\mathbb{Z}$. In other words, loops in
half-integer representations are screened to fundamental loops, and
loops in integer representations are screened to adjoint loops. In
the $SU(N)$ case we expect the same behavior \eq{Wasymptotics},
where now representations can differ by their $N$-ality. Concretely,
what we expect is that $E_\m$ is the string tension of a string in representation $\m$,
which is the totally antisymmetric representation of the same
$N$-ality as $\l$.

\subsection{$N$-ality of the eigenfuctions of the diffusion process}
To find the expectation value of the Wilson loop, we proceed as
follows. As before, the expectation value if computed by integrating
the loop operator over the diffusion ensemble
\be\label{Wloop}
\bra W_\l(\th)\ket =\int[\dd\th]\,W_\l(\th)\,p(t,\th)~,
\ee
where we are using the Haar integration measure (see Appendix \ref{group}).
$p(\th;t)$ is now the modified diffusion probability \eq{eqp}
expressed in the new set of eigenfunctions $\psi_\l$ satisfying the
Laplace equation \eq{qm} with the potential:
\be\label{laplace}
(\D-U(\th))\psi_\l(\th)=-E_\l\psi_\l(\th)~.
\ee
In this picture, the
$E_\l$'s are obviously positive definite because $U(\th)$ is (see Appendix
\ref{interaction}). From
now on we will often supress the $\th$-dependence, but this should
be understood in all the character formulas. We also need to express
the Wilson loop itself in the new basis. Thus, we define
\bea\label{Wexpand} W_\l(\th)&=&\chi_\l(\th)\nn
&=&\sum_\m\o_\l^\m\psi_\m(\th)~. \eea Finding the $\psi_\l$'s is now
equivalent to finding the coefficients $\o_\l^\m$.

There is a restriction on the eigenfunctions in which we expand the Wilson loop in \eq{Wexpand}. Remember that
the Wilson loop transforms as \eq{center} under center action. It is not hard to see that the
eigenfunctions must respect this property as well:
\be
\psi_\l(\th_i+{2\pi\over N})=e^{2\pi ik(\l)\over N}\psi_\l(\th_i)~.
\ee
This also means that the representations we sum over in \eq{Wexpand} all have
the same $N$-ality, in other words we sum over Young diagrams with
a fixed number of boxes modulo $N$. Thus we impose
\be
\o^\m_\n=0 ~~\mbox{if}~~k_\m\not=k_\n~.
\ee
Since this property is built into the definition of $\o^\m_\n$, we will not indicate this
restriction explicitly in the summation signs below.

\subsection{Solution of the diffusion equation: from Casimir to screening regime}

We now explicitly solve the Laplace equation \eq{laplace}. First of all we
invert \eq{Wexpand}:
\be
\label{psichi}
\psi_\l=\sum_\m c^\m_\l\chi_\m~.
\ee
Obviously, $c$ is nothing but the matrix inverse of $\o$:
\be
\sum_\n c^\n_\l\o^\m_\n=\sum_\n \o^\n_\l c^\m_\n=\d_\l^\m~.
\ee
Like before, $c^\m_\n$ vanishes whenever $k_\m\not=k_\n$.

It is now useful to expand the wave equation \eq{laplace} into
modes. Explicitly: \bea\label{expansion} \D\psi_\l&=&-\sum_\m
C_2(\m)c^\m_\l\chi_\m\nn U&=&\sum_\sigma u_\s\chi_\s\nn
U\psi_\l&=&\sum_{\m\n\s}c_\l^\m u_\s N_{\s\m}^\n\chi_\n \eea where
we used \be \chi_\m(\th)\chi_\l(\th)=\sum_\n
N^\n_{\m\l}\,\chi_\n(\th)~. \ee The $N$'s are the tensor
multiplicity coefficients for $SU(N)$, also called the
Littlewood-Richardson (for details, see Appendix \ref{interaction}).
Remember that $U(\th)$ has to preserve $N$-ality; therefore, \be
u_\s=0~~\mbox{if}~~k_\s\not=0~. \ee

Filling the above in \eq{laplace}, we get the following general
equation: \be \sum_\m c^\m_\l[(E_\l-C_2(\m))\chi_\m-\sum_\s\sum_\n
u_\s N^\n_{\s\m}\chi_\n]=0~. \ee Notice that this has now become a
purely algebraic equation. We can get a closed form for the
coefficients by multiplying the whole equation with
$\bra\cdot,\chi_{\m}\ket$ and using orthogonality of the characters.
We get: \be\label{solution} c^\m_\l(E_\l-C_2(\m))=\sum_{\m'} \sum_\s
c^{\m'}_\l u_\s N^\m_{\s\m'} \ee The Littlewood-Richardson
coefficients preserve $N$-ality, that is, $N^\m_{\s\n}=0$ if
$k_\m\not=k_\s+k_\n$. Therefore, since $k_\s=0$ and $c$ also
preserves $N$-ality, the above equation is trivial unless
$k_\m=k_\l$.

Taking $\m=\l$ in \eq{solution} and multiplying with $\o$, we get an
explicit expression for the energy:
\be
E_\l=\sum_{\m\n}\o^\l_\m c^\n_\l\left(\d^\m_\n C_2(\m)
+\sum_\s u_\s N^\m_{\s\n}\right)~.
\ee

At this point it is very convenient to introduce the following
notation:
\be
n^\m_\n=\sum_\s u_\s N^\m_{\s\n}
\ee
(for its properties, see Appendix \ref{interaction}).
Since $N^\m_{\s\n}$ preserves $N$-ality, it is clear that $n$ preserves it as
well:
\be
n^\m_\n=0~~if~~k_\m\not=k_\n~.
\ee
We define also the following bilinear form:
\be\label{g}
g^\m_\n=\d^\m_\n C_2(\m)+n^\m_\n~,
\ee
which by definition is $N$-ality preserving.
The energy can then simply be written as:
\be\label{energy}
E_\l=\sum_{\m\n}\o^\l_\m c^\n_\l g^\m_\n=(c^{-1}gc)^\l_\l~.
\ee
Going back to the full equation \eq{solution}, we can
rewrite it as
\be
E_\l\d^\m_\l=(c^{-1}gc)^\m_\l
\ee
and filling in the solution we found for the energy,
\be
(gc)^\m_\l=(c^{-1}gc)^\l_\l\,c^\m_\l~.
\ee
This  is an eigenvalue equation. Notice that
the index $\l$ labels the eigenvalues. Thus, regarding $c$ as a
vector, $c^\m_\l=c^\m_{(\l)}$, we get the eigenvalue problem
\be\label{eigenvaluepb}
g\cdot c_{(\l)}=(c^{-1}gc)_{(\l)}\,c_{(\l)}
\ee
where we now regard the $c$'s as eigenvectors of $g$. Thus, we
have reduced the problem of solving a differential equation on the
group, to the algebraic problem of finding eigenvalues of
eigenvectors of $g$, which is a given matrix once we choose the
potential, see \eq{g}. For simple potentials, $g$ is almost diagonal
and this equation can be readily solved.

We saw that the ground state of the system has zero energy. This implies:
\be
U(\th)={1\over\psi_0(\th)}\D\psi_0(\th)~.
\ee
Combining this with the expansion for $U$ \eq{expansion},
after some manipulations we get:
\be
\sum_\n c^\n_0 u_{\bar\n}=\sum_\n c_{0\n}
u_\n=0~,
\ee
where we used the charge conjugation properties of $c$ derived in Appendix
\ref{interaction}.

\subsection{Linear fluctuations}

For the purposes of this paper, it is enough to solve the diffusion equation
explicitly at the linearized level. Thus we can linearize equation \eq{solution}.
We set:
\bea
c^\m_\l&=&\d^\m_\l+\e\d c^\m_\l\nn
E_\l&=&C_2(\l)+\e\d E_\l\nn
u_\s&\rightarrow&u'_\s=\e u_\s
\eea
and work to first order in $\e$. Obviously, to lowest
order in $\e$ the equation is trivially satisfied: this is the
unperturbed solution. At the next order, we get two equations:
\bea\label{linen} \d E_\l&=&\sum_\s u_\s N^\l_{\s\l}\nn \d
E_\l\d^\m_\l+\d c^\m_\l(C_2(\l)-C_2(\m))&=&\sum_\s u_\l
N^\m_{\s\l}~. \eea This is easily solved: \be \d c^\m_\l={1\over
C_2(\l)-C_2(\m)} \sum_{\s\atop k_\s=0}u_\s N^\m_{\s\l}(1-\d^\m_\l)~.
\ee In other words, if $\hat n^\m_\n$ is the non-diagonal part of
$n^\m_\n$,
\bea\label{linsol}
c^\m_\n&=&\d^\m_\n -{1\over C_2(\m)-C_2(\n)}\,\hat n^\m_\n\nn
\o^\m_\n&=&\d^\m_\n+{1\over C_2(\m)-C_2(\n)}\,\hat n^\m_\n~.
\eea
Notice that the $\d c^\m_\m$
are not determined from the above. This is reflected by the fact
that the fluctuation is expressed in $\hat n^\m_\n$ not $n^\m_\n$.
This is due to the fact that we can always change them by
appropriate normalization. In particular, we can choose $\d
c^\l_\l=0$, in other words $c^\l_\l=1$ at least to second order in
perturbation theory. \\
\\
{\it Boundary conditions}\\
\\
We still need to discuss the choice of boundary condition \eq{conditions}.
It can be rewritten as:
\be
\sum_\l\chi_\l(\th)\,\mbox{dim}(\l)=\sum_\l\psi_\l(\th)\psi_\l(0)~.
\ee
Working this out, we get
\be\label{bc}
\mbox{dim}(\m)=\sum_{\l\n}c^\n_\l c^\m_\l\,\mbox{dim}(\n)~.
\ee
Taking $\m=0$, this reduces to
\be
\sum_{\l\n}c_\l^\n u_{\bar\l}\,\mbox{dim}(\n)=1~.
\ee
The full boundary condition equation can be rewritten as
\be
\sum_\l(c_{\bar\m}^{\bar\l}-c^\m_\l)\,\mbox{dim}(\l)=0~.
\ee

We can now analyze these boundary conditions at the linear level. We get:
\be
\sum_\l{1\over C_2(\m)-C_2(\l)}\left(\mbox{dim}(\l)\,\hat n^\m_\l-\mbox{dim}(\m)\,\hat n^\l_\m\right)=0~.
\ee
When $\m=0$, we get the following condition on the potential:
\be
\sum_\l{1\over C_2(\l)}\,(\mbox{dim}(\l)-1)\,u_\l=0~.
\ee

\subsection{Extracting the $k$-strings}

We have now solved the model in the previous section in implicit form,
the solution being given in equations \eq{energy} and
\eq{eigenvaluepb}. For linear perturbations, we obtained a closed form for the solution in
\eq{linen} and \eq{linsol}. We are now ready to compute the expectation
value of the Wilson loop and extract the tension of the minimal string.

Before going into details, let us briefly describe what will happen.
The important point is the following: at
large $t$, because of the exponential damping, the leading
contribution will be the one with lowest energy $E_\m$. We are summing
over all representations $\m$ that
have the same $N$-ality $k_\l$ as the loop operator $W_\l$. The
leading contribution is therefore given by the representation $\m$ that minimizes
$E_\m$. Since the leading term is
\be\label{em}
E_\m=C_2(\m)+{\cal O}(\e)~,
\ee
we see that the leading contribution to a Wilson loop $\bra W_\l\ket$
will be given by the representation of lowest Casimir. But that is
the totally antisymmetric representation with $N$-ality $k_\l$. Thus
we have found the $k$-string! Thus, we get
that a Wilson loop in representation $\l$ is screeened to a loop in
representation $\m$, where $\m$ is the totally antisymmetric
representation with $k_\l$ quarks. Of course in the $SU(2)$ case we
reproduce the result of \cite{blnt}, but we see that in general this
model is very powerful in predicting the screening behavior of the
background. This conclusion remains unchanged taking into account small perturbations
in \eq{em}, and in general any perturbation produced by the potential,
as long it keeps the ordering of the lowest-lying energy levels.

We now give the details of the above. All the ingredients to compute the
Wilson loop average have been given earlier:
\bea\label{endresult}
\bra W_\l(\th)\ket&=&\int[\dd\th]\,
W_\l(\th)\,p(t,\th)=\sum_{\m\n}{\psi_\m(0)\over\psi_0(0)}\,\o^\n_\l e^{-E_\m t}
\int[\dd\th]\,\psi_\m(\th)\psi_\n(\th)\psi_0(\th)\nn
&=&\sum_\m{\psi_\m(0)\over\psi_0(0)}\,\o^{\bar\m}_\l\,e^{-E_\m t}~.
\eea
Notice that in this computation it was important to use the fact that $p(t,\th)$,
rather than the correlation function $g(t,\th)$, is the correct diffusion
density; one indeed easily checks that it is normalized to one,
\be
\int[\dd\th]\,p(t,\th)=\sum_\l{\psi_\l(0)\over\psi_0(0)}\,e^{-E_\l t}\int[\dd\th]\,\psi_\l(\th)\psi_0(\th)=1~.
\ee

Let us now look at what happens for large $t$:
\be\label{asympt}
\bra W_\l(\th)\ket={\psi_{\sm{A}(\l)}(0)\over\psi_0(0)}\,\o^{\sm{A}(\l)}_\l\,e^{-E_{\sm{A}(\l)}t}
\ee
where A($\l$) means the antisymmetric representation of
$N$-ality $k_\l$. In the case that $\l$ is in the same $N$-ality
class as the trivial representation, the loop gets screened to:
\be
\bra W_\l(\th)\ket=\o^0_\l~,
\ee
that is, a constant.

The above computations were exact. We now work out the linearized
case explicitly. We get:
\be
g(t;\th,\th')=\sum_\l\chi_\l(\th)\chi_\l(\th')e^{-(C_2(\l)+n^\l_\l)t}+\sum_{\l\l'}{\chi_\l(\th)\chi_{\l'}(\th')
\over C_2(\l)-C_2(\l')} (\hat n^{\l'}_\l\, e^{-C_2(\l)t} -\hat
n^\l_{\l'}\,e^{-C_2(\l')t})~.
\ee
The additional term correctly respects the symmetry in $\th$ and $\th'$, as it should.

\subsection{Specific choices of the potential: examples}\label{examples}

The approach of \cite{blnt} is phenomenological, and extra input is required to determine a phenomenologically intersting
potential. For illustration purposes, and since the model with the potential is quite interesting in its own right, we now
give some examples of simple potentials. A first rather trivial example is where $u$ is only non-vanishing for the trivial
representation:
\be
u_\s=u\,\d_{\s,0}
\ee
where $u$ is a number. We get
\be
n^\m_\n&=&u\,\d^\m_\n\nn
c^\m_\n&=&\d^\m_\n\nn
E_\l&=&C_2(\l)+u~.
\eea
Thus in this case we see that $c$ remains unmodified, and the whole effect of the potential is to shift the zero-point
energy of the theory. From the point of view of 2d Yang-Mills, this is adding a cosmological constant term to the action.

We now look at the following less trivial example:
\be
u_\s=u\,\chi_{\bar\s}(\phi)~,
\ee
the $SU(N)$ character for some {\it fixed} $\phi$, that is, a collection of numbers, and $u$ is again an overall constant.
There are two limiting cases:
\bea
\chi_\m(0)&=&\mbox{dim}(\m)\nn
\chi_\m(\r)&=&\mbox{dim}_q(\m)
\eea
where in the second case we have conjugated the group element $\f$ to the trivial representation $\r$ by exponentiation
with $q$. We get:
\be\label{qdim}
n^\m_\n=u\,\chi_{\bar\m}(\f)\chi_\n(\f)~.
\ee
This result is exact. Linearizing, we get:
\bea
\d E_\l&=&u\,(\chi_\l(\f))^2\nn
\d c^\m_\l&=&-{u\over C_2(\m)-C_2(\l)}\,\chi_{\bar\m}(\f)\chi_\l(\f)~,
\eea
where $\m\not=\n$. The expectation value of the Wilson loop is now:
\be
\bra W_\l(\th)\ket = {\psi_{\bar\l}(0)\over\psi_0(0)}\,e^{-E_{\bar\l}t} -u\,{\chi_\l(\phi)\over\chi_0(0)}
\sum_\m{\chi_\m(\phi)\chi_\m(0)\over C_2(\l)-C_2(\m)}\,e^{-C_2(\m) t}~.
\ee
Notice that the correction to the energy in this simple example is the square of the quantum dimension of the
representation $\l$. Asymptotically this leads to the quantum dimension of the antisymmetric representation of
$N$-ality $\l$, which is somewhat reminiscent of the sine law. We will get back to this point in the conclusions.

\section{Wilson loop effective action and small time limit}

\subsection{A proposal for an effective Wilson loop action}

As recalled in the introduction it is quite hard to study the Wilson
loop dynamics. One of the outcomes of the approach we are following
is that one can write an effective action for the Wilson loop
dynamics. The $SU(2)$ case was already treated in \cite{blnt}. Here
we revisit the problem and sketch the $SU(N)$ generalization.

Before doing this we would like to recall that the mapping from the
description with a drift term to the one with the potential can be
obtained in a compact and elegant way as follows: start from the the
operator $\tilde\D$ in (\ref{tiD}). To go to a hamiltonian
description with a potential perform a similarity transformation \be
H= T \tilde\D T^{-1} \ee with the operator $T$ given by $T = \exp
(V/2)$. One can easily check that the hamiltonian $H$ will be given
by $\D -U$, with $U$ precisely given by (\ref{potentialU}). Note
that (up to normalization constants) if one defines \be C_{\theta_i}
= \partial_{\theta_i} +\frac{1}{2}
\partial_{\theta_i} V ; \hspace{1cm} C_{\theta_i}^+ = -
\partial_{\theta_i} +\frac{1}{2} \partial_{\theta_i} V
 \ee
then the hamiltonian can be written as \be H= \sum_i C_{\theta_i}^+
C_{\theta_i} \ee This allows to prove more easily that $H$ is a
self-adjoint operator with real eigenvalues. In addition the ground
state $\chi_0$ is given by $C_{\theta_i} \chi_0 =0 $ which once
solved gives precisely $\exp (-V/2)$. This is the corresponding
infinite time (i.e. equilibrium state) in the Fokker-Planck language
\cite{Zinn-Justin:2002ru}.

Consider now the effective action. In the $SU(2)$ case the
\cite{blnt} proposal was to write the probability distribution as
\be\label{pathi}
P(t',t,\theta',\theta) =
\int_{\theta(t')=\theta'}^{\theta(t)=\theta}[\dd\th(t)]\, e^{-S_{\sm{eff}}[\theta(t)]}~.
\ee 
We went back to the probability distribution defined in \eq{P}. This
is related to the two-point function $g(t',t,\th',\th)$ by the Jacobian.
Tthe effective action is given by \be \label{effsu2}
S_{\sm{eff}}[\theta(t)]=\int\dd t\left({1\over4}\dot
\theta^2(t)+U(\theta (t))\right)~. \ee

Let us review briefly before moving to $SU(N)$ some notions from the
theory of stochastic differential equations
\cite{Zinn-Justin:2002ru}. Start from Langevin equation\footnote{Here
we assume $i=1,..,N-1$ as in our set up and the
$\theta_i (t)$ are then the  coordinates in the Langevin
description. They have -{\it no}- relation with the $\theta_i$ (time
independent, i.e. just coordinates)appearing in the Fokker-Planck
equation we have been considering. The coordinates in the effective
action (\ref{effsu2})(which are time dependent indeed) are thus the
one in the Langevin description. We will always put explicitly the
time dependence when we consider the $\theta_i (t)$ entering in the
Langevin description.}
\be
\dot\theta_i = -\frac{1}{2}
f_i(\theta(t)) +\eta_i (t)
\ee
where $f_i$ is a function of ${\bf \theta}(t)$ and $\eta_i$ the noise
(which we assume Gaussian). Then
the probability $g(\theta,t)$ which can be show
\cite{Zinn-Justin:2002ru} to satisfy the Fokker-Planck equation can
be expressed as
\be
P(\theta,t)= \bra \prod_{i}^{N-1} \delta [\theta_i
(t)- \theta_i] \ket_\eta~,
\ee
where the average is taken in
this case with respect to the gaussian noise.

The effective action appearing in \eq{pathi} is given by: 
\be \label{effstoch} 
S_{\sm{eff}}[\theta(t)]=
\int_{t'}^{t''} \dd t\, \frac{1}{2} [(\dot\theta_i (t)+
\frac{1}{2} f_i(\theta (t))^2 - \Omega \frac{1}{2} \partial_i f_i
(\theta (t)) ]~. \ee

This action, once one assumes $f_i = \Omega \partial_i V$ where ($\Omega$
is the diffusion coefficient which appears in the two point function
of the gaussian noise), i.e. a purely dissipative Langevin equation
which will produce a drift term of the kind we are using, can be
checked to correspond to the action (\ref{effsu2}) proposed in
\cite{blnt} for $SU(2)$ where of course we keep only one $\theta$.

On a general Riemannian manifold, however, the probability
\footnote{Here Stratanovich convention is assumed for the operator
ordering.}
\be
P(\theta'',t'';\th',t') = \int_{\theta(t')=
\theta'}^{\theta(t'')=\theta''}  \dd \theta (t) \prod_i [\det (e)]^{-1}
\exp (-S_{\sm{eff}})
\ee
where now the action has a complicated expression
\cite{Zinn-Justin:2002ru} containing vielbeins $e_\mu^a$ and their
derivatives. It would be interesting to generalize the study to $SU(N)$ case. 

In addition in the computation above we always set the diffusion
coefficient $\Omega=1$. But in this effective action approach it
plays the role of $\hbar$. Restoring it means that some terms will be
dominant with respect to others. For instance the second term on the right-hand
side of (\ref{effstoch}) can be shown to be subleading. This is relevant
if one has in mind to do saddle points approximations around minima
of the potential. If one kept indeed only the leading order terms
one would not be able to resolve as usual the degeneracy of the extrema.

\subsection{Small time limit: diffusion on the Lie algebra}

Consider the quantity $\tau (\rho)$, namely the variation of the
size of the loop $\rho$ with respect to time $t$.\footnote{The size
of the loop $\rho$ has not to be confused of course with the
spectral density $\rho (\theta,t)$ which we have been discussing and
in our case is given by $G(\theta,t)$. We keep this notation to make
contact with \cite{blnt}.} To get some more explicit information
about this function, we will follow \cite{blnt} and look at two particular limits of the
diffusion probability. From the kinetic term we get
\be
\int\dd\r\,\t(\r)\left(\dd\th_i\over\dd\r\right)^2~,
\ee
where 
\be
\tau(\r)=\left({\dd t\over\dd\r}\right)^{-1}~.
\ee
So we look in the regime where the
potential becomes irrelevant and we can use the free diffusion
problem: \be\label{freediffusion}
p(t,\th)=\sum_\l\chi_\l(\th)\,\mbox{dim}(\l)\,e^{-C_2(\l)t}~. 
\ee 
Since we are considering Brownian motion on the circle \cite{sdh1,sdhmt},
we can use Poisson resummation to rewrite this as \cite{mo}
\be\label{kernel} 
p(t,\th)={1\over(2\pi t)^{N^2-1\over2}}\sum_{l_i}{D(\th_i+2\pi l_i)\over J(\th_i+2\pi
l_i)}\,e^{-\sum_i(\th_i+2\pi l_i)^2/2t}~.
\ee 
See Appendix \ref{kernels} for full details and the definitions of
$D$ and $J$.

In the $SU(2)$ case, this reads explicitly:
\be
p(t,\th)={1\over(2\pi t)^{3/2}}\sum_{n=-\infty}^\infty{\th+2\pi n\over\sin\th}\,e^{-(\th+2\pi n)^2/t}~.
\ee
It is now possible to take the limit $t\rightarrow0$ directly. The only term contributing gives:
\be
p(t,\th)={1\over(2\pi t)^{3/2}}\,e^{-\th^2/t}~,
\ee
which is the result in \cite{blnt}, and we are also taking the $\th$'s to be small. In doing so, we have broken the
periodicity of $\th$. This method immediately generalizes to $SU(N)$. Taking $t$ small in \eq{kernel}, we get
\be
p(t,\th)\simeq{1\over(2\pi t)^{N^2-1\over2}}\,{D(\th_i)\over J(\th_i)}\,e^{-\sum_{i=1}^N\th_i^2/2t}
={1\over(2\pi t)^{N^2-1\over2}}\,
e^{-\left({1\over2t}-{N\over24}\right)\,\sum_{i=1}^N\th_i^2}~.
\ee
The factor of $N\over24$ comes from the measure factor $D/J$, but for the limit of small $t$ and finite $N$ that we are considering for
the moment we can drop this factor. Thus the result is that the small $t$ limit of the diffusion probability is the kernel
(see Appendix \ref{kernels})
\be
p(t,\th)\simeq{\cal K}(t,H)~,
\ee
which is the fundamental solution of the heat equation on the full Lie algebra $su(N)$. But as explained in Appendix \ref{kernels}
this is the free Brownian motion probability on a space of dimension $N^2-1$:
\be\label{measure}
{\cal K}(t,\th_i)\,[\dd H]=k_{N^2-1}(t,\th_i)\,\dd\th_1\ldots\dd\th_{N^2-1}=\prod_{i=1}^{N^2-1}p_{U(1)}(t,\th_1)\dd\th_i~.
\ee
Thus we recover the free Brownian motion on the tangent space of the group, which is what we expect.
The second form in \eq{measure} is the product of the diffusion probabilities of $N^2-1$ abelian $U(1)$ theories. The Wilson loop distribution
in that theory was computed in \cite{blnt}:
\be
p_{U(1)}(t,\th)={1\over\sqrt{4\pi\k}}\,e^{-{\th^2\over4\pi\k}}
\ee
where
\be
\k={g^2\over4\sqrt{2\pi}}{\r\over\e}~,
\ee
and $\e$ is the smearing of the loop, which acts as a regulator. Thus, we get that
\be\label{limit1}
t=\sqrt{\pi\over2}\,{g^2\r\over\e}~,~\r\rightarrow0~.
\ee
The effective action in this regime is thus
\be
S_{\sm{eff}}[\th_i(t)]=\sqrt{2\over\pi}{\e\over g^2}\int\dd\r\,{1\over4}\left(\dd\th_i(t)\over\dd\r\right)^2~.
\ee

Let us comment on the limit $t\rightarrow0$ that we took above. It
was useful to do the Poisson resummation \eq{kernel} because it
gives us an expression where we only need to keep a single term in
this limit. If nevertheless we wanted to work directly with
\eq{freediffusion}, we could have replaced the sum over
representations by an integral. This is readily done by rescaling
$\l_i$ by $\sqrt{t}$ in the usual way, so we
consider large values of $\l_i$ and keep $\l_i'=\l_i\sqrt{t}$ fixed.
Notice that in doing so we are not taking any large $N$ limit.
The
integral one then gets is the usual Hermitian matrix model with
Vandermonde interaction. Alternately, one can deform the theory by
considering $p(t;\th,\th')$ with non-vanishing $\th'$. One can then
apply the techniques of \cite{sdhmt2} to get a Stieltjes-Wigert
matrix model. At the end one can take the $\th'\rightarrow0$
limit.


\subsection{Casimir scaling regime}

Next we look at the Casimir scaling regime for large loops, where
the potential is still irrelevant and we have an area law. The
Casimir in the fundamental representation $\tableau{1}$ is
$C_2(\tableau{1})=N-1/N$, so we find that \be\label{limit2}
t={\s\over N-1/N}\,\pi\r^2~. \ee for large $\r$. Thus, assuming that
like in \cite{blnt} that the logarithm of the Wilson loop is a sum
of a constant, perimeter and area terms, the shape of the function
$\tau(\r)$ that respects the two limits \eq{limit1} and \eq{limit2}
is \be \tau(\r)=\left(\sqrt{\pi\over2}{g^2\over\e}+{2\pi\s \rho
\over N-1/N}\right)^{-1}~. \ee

\section{Vortex density and links with Lattice Gauge theory}
Once the spectral density is known one can also consider the simple
vortex density; this is a quite interesting quantity to compute
considering the importance of the center symmetry in the confinement
problem and for our set up as well. In
\cite{Belova:1983rd,Makeenko:1982gr} the $SU(2)$ case for
Wilson loops in the fundamental representation was addressed. For
simplicity we consider here the same situation.

Recall that fluctuations of the gauge group center, namely $Z_2$ in
this case, are expected to be suppressed as soon as one moves to
weak coupling but becomes pretty relevant at strong coupling. In
particular various Monte Carlo simulations \cite{Makeenko:1982gr}
show an almost exponential fall off (we are going to
comment on this just below) for $\beta>2$ of the thin vortex density
$\bar{E}$. The latter is defined as follows
\be
\bar{E}= \frac{1}{2} \left(  1- \bra\mbox{sign}(W_j)\ket   \right)
\ee
where $j=1/2$ of course for the fundamental rep we are interested. In our
case then $W_{j=1/2} = \cos\theta$ and one easily gets
\be
\bar{E}=\int_{\pi/2}^{\pi} \dd \theta \,\sin^2 \theta\, G(\theta,t)
\ee

Inserting our expression for $G(\theta,t)$ one gets
\be \bar{E}= \frac{1}{2} + \frac{2}{\pi} \sum_{n=2}^\infty
\frac{n^2}{n^2-1} e^{-(n^2-1)t} \cos(n \pi /2) \ee

The first term in our results is exactly the same found in
\cite{Makeenko:1982gr}. We get qualitative agreement for the plot of
$\bar{E}$ vs $\beta$., i.e. the density decreases as soon as $\beta$
grows as expected. It is quite difficult however to make a
quantitative agreement for the following reasons which have to do
with dimensional reduction. First of all, the four dimensional
$\beta$ and the two dimensional $\beta_{2d}$ are not the same but
schematically $\beta= b \beta_{2d}$ as said. For small $\beta$, $b
\rightarrow 1$ but for large $\beta$ then $b \rightarrow 2$. This,
as explained, is the way in which dimensional reduction works at the
level of the couplings. Therefore in the Monte Carlo simulations one
can only fit the data -after- choosing $b$ in a suitable way. In our
case we have an additional issue. Namely, the time $t$ is inversely
proportional to $\beta$ but also depends on the size of the loop.
This is additional residuum of the four dimensional physics in the
dimensional reduction scenario. Therefore it sounds quite hard to
provide quantitative matching with Monte Carlo data, but at
qualitative level our model provides, consistently, the same leading
order value at strong coupling and a progressive fall off as soon as
one goes to weak coupling.\\
\\
{\it Links with lattice gauge theory}\\
\\
We believe that the lattice gauge theory and in particular the
geometrical LGT approach can give at this stage a complementary and
corroborative perspective to the whole scenario. This is also useful
to explain a little better the meaning of the time $t$ entering in
the diffusion process.

 Geometrical LGT actions indeed (\cite{mo} and
references therein) are supposed to describe fluctuations of
plaquettes in the weak coupling regime; they are indeed weak
coupling approximations of the standard LGT Wilson action. The
prototype is the abelian Villan action \cite{Villain:1974ir} for
magnets which is eventually generalized to a non abelian group as
\be \label{geomaction} \exp ( S_J) = \sum_r d_r \chi_r(U) e^{-C_r/(N
\beta_J)} \ee where the sum is over all irreducible unitary
representations of dimension $d_r$; $\chi_r$ is the character of the
plaquette $U$, $C_r$ the (quadratic) Casimir in the representation r
and N depends on the group we are considering. This action clearly
satisfies the heat equation on the group manifold  with initial
condition $\delta(U,1)=\exp (S_J(U,1))$. From physical point of view
we imagine to study small (i.e. quadratic fluctuations) of the
plaquette U around unity in group space; we use the geodesic
distance of the plaquette from unity to measure them and the metric
is induced on the group manifold by the invariant quadratic form of
the Lie algebra.

 ``Time" has therefore to be interpreted as the
response to the heating and goes like $1/\beta$. Therefore the large
time behavior we have been discussions means strong coupling regime
and the initial conditions that one is starting from weak coupling
to measure heat propagating.

However, after we define the geometrical action we have to put it on
a specific d-dimensional lattice configurations as usual in LGT to
-concretely- compute. Therefore the total action will actually be
the product of the single action one for each plaquette and the Haar
measure entering in a path integral description will carry link
indices. As usual it is clearly not possible to do an exact
computation. If on the other hand one invokes the dimensional
reduction hypothesis the geometrical action becomes precisely only
at this stage the one of two dimensional QCD$_2$ which is a character
expansion indeed and one can forget now about all the plaquette
links indices.\footnote{ As an aside remark, note that the
dimensional reduction is supposed to work not only for $d=4
\rightarrow d=2$ but for every $d \rightarrow d=2$.
\cite{Ambjorn:1984mb}.}

We would like to make a simple example with the LGT geometrical
action to reproduce the result of \cite{blnt}. Consider free $SU(2)$ for
simplicity. The vev of the wilson loop in the spin j representaion
is
\be \label{Wvev}
\bra W_j\ket = \int [DU] e^S W_j~.
\ee
The geometrical action can be easily computed and becomes in the
$SU(2)$ case
\be
 e^S=
\sum_{2j=0}^{\infty} (2j+1) \frac{\sin((2j+1) \theta)}{\sin(\theta)}
\exp(- \frac{j(j+1)}{2 \beta}) \ee
 and \be W_j= \frac{1}{(2j+1)} Tr_j U \ee coincides with the $SU(2)$ character in the spin $j$
 representation. Inserting into (\ref{Wvev}) one gets
\be
\bra W_j\ket = \exp ( -\frac{j(j+1)}{2 \beta})
\ee
which is precisely
what obtained in \cite{blnt} at the qualitative level in the free diffusion case.


\section{Discussion and conclusions}

Building on the results of \cite{blnt}, we have presented a phenomenological
model that describes the transition from Casimir scaling to screening. Motion on
the group manifold $SU(N)$ with an arbitrary potential breaks the simple Casimir
scaling. The assumption of center symmetry of the potential predicts that the
stable strings are the $k$-strings. We have shown that asymptotically, the tension of the string
does not depend on the representation of the Wilson loop, but only on its
$N$-ality class. For intermediate areas, the expectation value of the Wilson is a
sum over all possible representations of the same $N$-ality, and this involves
non-trivial dynamics. At large $N$, we expect the expectation value of the Wilson loop
to be given by a two-matrix model. We will get back to this in the future \cite{adhg}. It would
also be interesting to see if phase transitions of the Douglas-Kazakov type are robust under
perturbation by an arbitrary potential.

A generalization to include time-dependent potentials as suggested in \cite{blnt} should be
straightforward. One can then use time-dependent perturbation theory as in quantum mechanics. Details
of this will appear elsewhere \cite{adhg}. Our methods can also be applied when the potential no longer
respects center symmetry; this is the case if dynamical quarks are introduced \cite{blnt}.

Our model is consistent with a closed string expansion in powers of $N$. Namely, it
has an expansion in even powers of $1/N$ like the large $N$ expansion of QCD$_2$ \cite{gross,gt}.
We want to stress however that our model is {\it not} QCD$_2$. QCD$_2$ is asymptotically trivial
for large areas; the partition function goes to one, and there is no dynamics for large Wilson
loops. Our theory however is non-trivial in this limit, see \eq{asympt}: the leading contribution
to the Wilson loop is still interacting.

One important question is what is the actual form of the potential? The simple symmetry
requirement \eq{sympot} was enough to obtain the right asymptotic behavior. To say more about the
actual form of the potential, however, more phenomenological input is required. We leave this for
the furure. Here we have shown that the model with general potential is
solvable in group theoretic terms once the potential is provided. This is an interesting model to
study in its own right. Providing the potential amounts to giving a finite set of numbers $u_\s$.
The model in principle contains enough freedom to accomodate for various values of the string tension.
The only actual restriction is that the tension depends only on the $N$-ality class of the representation,
and its leading contribution is given by the Casimir of the antisymmetric representation of that $N$-ality.
Because of this freedom, this model might in principle be able to mimic other kinds of behavior adequate to
for example supersymmetric models \cite{shifman,as}, where the string tension is given by the sine-formula:
\be\label{sine}
\s_k=N\L^2\sin\left({\pi k\over N}\right)~,
\ee
rather than the Casimir formula. As argued in \cite{as}, both formulas give the same result as $N\rightarrow\infty$,
namely $k$ times the tension of the fundamental string (of $N$-ality one):
\be
\s_k=k\s_1~.
\ee
Thus, the difference between both formulas is of order $1/N$, and in principle this could be taken as the
parameter that measures the perturbation of our potential. Thus, the phenomenological potential of \cite{blnt}
could in principle interpolate between both models. In fact, in one of the examples considered in Section (\ref{examples}) the
correction to the ground-state energy came out to be the square of the quantum dimension, which applied to the tension of the
stable string gives the quantum dimension of the antisymmetric string. This is already very reminiscent of the sine formula.
However, we should remember that this is only a correction above the ground state energy.
A more promising approach seems the following.
Notice that both the sine law and Casimir scaling have invariance under $k\leftrightarrow N-k$.
If we view $k$ as the level of some conformal field theory, this is level-rank duality. A second remark is that \eq{sine}
is the quantum-deformed level, where the $q$-deformation parameter is $q=e^{2\pi i\over N}$. Thus, it is tempting to speculate
that the sine formula is
somehow related to a $q$-deformation. In fact, if we considered motion of the particle on the quantum group manifold,
repeating the steps in this paper would replace the Casimir by the quantum Casimir (see for example \cite{chugoddard}).

There are many points that deserve further research. First of all, the expectation value of the Wilson loop computed in
\eq{endresult} itself, and not only the probability distribution, satisfies a heat equation. Presumably this heat equation
can be seen as a loop equation. It would be interesting to analyze this in detail.

It would be interesting to see if, apart from the phenomenological and mathematical interest of the model in \cite{blnt}, it
can teach us something about dimensional reduction and the underlying theory describing screening. It is likely that the
potential can be seen as an effective action
that arises after integrating out some degrees of freedom. These degrees of freedom could be matter in an $SU(N)$ representation
of zero $N$-ality like the adjoint representation. Also, if we think of this matter as interacting with the gluons in a
dimensionally reduced theory, in the screening phase the background necessarily will have to break the invariance under area-preserving
diffeomorphisms. Simple cases of such models have been considered in \cite{gkms}. It would be
interesting to pursue this further.

In addition it would be interesting to explore more in detail the
Wilson loop effective action in the $SU(N)$ along the lines sketched
in Section 6 and in particular study in detail the large $N$ limit of
the model.

Finally one could try to make a link along the lines of \cite{sdh2,sdhmt,D'Adda:2001nf} 
with vicious random walk models trying to make a connection with the results of
\cite{katori}. We leave all these topics for future research \cite{adhg}.

\section*{Acknowledgements}
We thank A.~Armoni, A.~Petkou, and especially S.~Eliztur for many
useful discussions. The research of GA has been supported by a Marie
Curie Fellowship under contract MEIF-CT-2003-502412, partially by a Golda Meir
Fellowship, the ISF Israel Academy of Sciences and Humanities, the
GIF German-Israel Binational Science Found and the European Network
RTN HPRN-CT-2000-00122. SdH and PG thank the Third Simons Workshop
on Mathematics and Physics, where part of this collaboration
started. GA thanks AEI-Golm for hospitality during various stages of
this work.

\appendix
\section{Laplacians and heat kernels on group manifolds}\label{kernels}
In this Appendix we discuss the heat equation on the group manifold
and the heat equation on the algebra(i.e. the full algebra and its
Cartan subalgebra) we considered in the in the paper. The group
manifold is in our case $SU(N)$, namely the color group of the gauge theory.

We first recall the expression for the radial Laplacian $\D$ on
$SU(N)$ which we considered in the paper 
\bea 
\D&=&\half{1\over J^2(\theta)}{\pa\over\pa\theta_i}J^2(\theta){\pa\over\pa\theta_i}\nn
&=&{1\over2J(\theta)}\,{\pa^2\over\pa\theta_i^2}
J(\theta)+{1\over24}\,N(N^2-1)\nn 
J(\theta)&=&\prod_{i<j}2\sin\half(\theta_i-\theta_j)\nn 
D(\th_i)&=&\prod_{i<j}(\th_i-\th_j)\nn~.
\eea 
where $J(\theta)$ and $D(\theta)$ are Jacobian the Vandermonde
determinant of the Hermitian and unitary $U(N)$ matrices. To get
$SU(N)$ one simplify the $N$ variables $\theta_i$ add up to zero:
$\sum_{i=1}^N\theta_i=0$.

Note the term proportional to $N(N-1)$ is just the curvature of the
manifold and being a constant it can be shifted away. As noticed
before we consider only class functions therefore we restrict to the
radial Laplacian. For a rigorous derivation of the full Laplacian
including the angular part see \cite{helgason} and also \cite{to}
for a simplified treatment. We report in any case the expression of
the angular part which is given by
 \be
\label{angularlaplacian}  - \sum_{i<j} \frac{L_{ij}^2 + M_{ij}^2}{16
\sin^2[(\theta_i -\theta_j)/2]} \ee with $i L_{ij} = E_{ij}- E_{ji}$
and $iM_{ij}= i(E_{ij}+E_{ji})$, where the $E_{ij}$ matrices are
zero everywhere except one at the $(i,j)$ entries. The diagonal part
is spanned by the $N-1$ Cartans while the $L_{ij}$ and $M_{ij}$ are
sort of generalized $N^2-N$ step operators: in the simple $SU(2)$
case they are just the usual $J_+$ and $J_-$.

We now recall some useful results from \cite{mo} concerning heat
kernels. The fundamental solution of the heat equation \be
\left(\pa_t-\D\right)K(t,\theta)=0 \ee where $\D$ is the group
Laplacian, is given by \footnote{Recall that in the $SU(N)$ case
$\sum_{i=0}^{N} l_i =0.$} 
\be
K(t,\theta)=\sum_\l\chi_\l(\theta)\chi_\l(0)\,e^{-tC_2(\l)}={1\over(2\pi
t)^{N^2-1\over2}}\sum_{l_i}{D(\theta_i+2\pi l_i)\over J(\theta_i+2\pi
l_i)}\,e^{-{1\over2t}(\f_i+2\pi l_i)^2}~. \ee We also consider the
solution \be K_0(t,\theta)={1\over(2\pi
t)^{N^2-1\over2}}{D(\theta_i)\over
J(\theta_i)}\,e^{-{1\over2t}\sum_i\theta_i^2}~. \ee which is not,
however, periodic \cite{Dowker1} in $\theta$. The relation between
both is \be K(t,\theta_i)=\sum_{l_i}K_0(\theta_i+2\pi l_i)~. \ee
Thus, we see that $K_0$ corresponds to the kernel in the covering
space. In that
sense, $K_0$ lives in the {\it algebra} whereas $K$ lives on the {\it group}.

At small $t$, they are identical: \be \lim_{t\rightarrow
0}K(t,\f)=K_0(t,\f)~. \ee

The $t$-dependence above suggests that this solution is related to
the solution of the heat equation on the space of Hermitian
matrices: \be \left(\pa_t-\D_H\right){\cal K}(t,H)=0 \ee where \bea
{\cal K}(t,H)&=&{1\over(2\pi t)^{N^2/2}}\,e^{-{1\over2t}\,\sm{Tr}
H^2}\nn &=&{1\over(2\pi
t)^{N^2/2}}\,e^{-{1\over2t}\sum_{i=1}^N\theta_i^2}~. \eea In the
first form, $H$ is a matrix and so the trace depends $N^2-1$ of its
entries. In the second line we have done a similarity transformation
$=S\,\mbox{diag}(\theta_1,\ldots,\theta_N)S^{\dagger}$ with a
unitary matrix $S$. Notice that if $U=\exp(iH)$ is a unitary matrix,
$S$ is the same transformation that diagonalizes $U$:
$U=S\,\mbox{diag}(e^{i\theta_i},\ldots,e^{i\theta_N})S^{\dagger}$.

It is now clear how both kernels are related: \be
K_0(t,\theta)={D(\theta)\over J(\theta)}\,{\cal K}(t,\theta) \ee and
\be K(t,\theta_i)=\sum_{l_i}{D(\theta_i+2\pi l_i)\over
J(\theta_i+2\pi l_i)}\,{\cal K}(t,\theta_i)~. \ee At small $t$, \be
K(\theta_i)\simeq {D(\theta_i)\over J(\theta_i)}\,{\cal
K}(\theta_i)=K_0(\theta_i)~. \ee

Finally, we define the Lie algebra kernel \be
k_{N-1}(t,\theta)={1\over(2\pi
t)^{N-1\over2}}\,e^{-{1\over2t}\sum_{i=1}^N \theta_i^2} \ee which
satisfies the heat equation on the Cartan subalgebra of $SU(N)$: \be
\left(\pa_t-\D_{\sm{C}}\right)k_{N-1}(t,\theta)=0~. \ee The above is
just the product of $N$ one-dimensional Brownian motions.

The kernels are correctly integrated as follows: \bea
1&=&\int_{\mathbb{R}^{N^2-1}}[\dd H]\,{\cal
K}(t,H)=\int_{\mathbb{R}}\dd\varphi\,\varphi^{N^2-2}{\cal
K}(t,\varphi)
=\int_{\mathbb{R}^{N-1}}\dd\theta_1\ldots\dd\theta_N\,k_{N-1}(t,\theta)\nn
1&=&\int\dd\th\,K(t,\th)=(2\pi t)^{N^2-N\over2}\int[\dd U]K(t,U)
\eea $[\dd H]$ is as usual the Hermitian matrix measure, $\dd\th$
the diagonal part of the unitary measure,
$\dd\th=\dd\th_1\ldots\dd\th_NJ(\th_i)$, and $[\dd U]$ the unitary
measure $[\dd U]=[\dd S]\,\dd\th$ where $S$ is the similarity
transformation used to diagonalize $U$. $\theta$ denotes a radial
coordinate for the $N^2-1$ variables of $H$. In particular, as
densities we have \be {\cal
K}(t,\vf)\,\vf^{N^2-1}\,\dd\vf=k_{N-1}(t,\theta)\,\dd\theta_1\ldots\theta_N~,
\ee which is what we used in the text.



\section{Some elements of group theory }\label{group}

We review some group theory notions which are used throughout the
paper following the notations and conventions of
\cite{DiFrancesco:1997nk}.

A weight written in the fundamental weight basis is: 
\be
\l=\sum_{i=1}^N\l_i\o_i=\sum_{i=1}^N(\ell_i-\k)\e_i \ee where $\e_i$
are unit vectors in $\mathbb{R}^N$ and \be \k={1\over
N}\sum_{j=1}^{N-1}j\l_j~. \ee So $k=\k N\,(\mbox{mod}\,N$) is the
$N$-ality. The relation between both bases is: \bea
\ell_i&=&\sum_{j=i}^{N-1}\l_j\nn \o_i&=&\sum_{j=1}^i\e_i-{i\over
N}\sum_{j=1}^N\e_i \eea for $i=1,\ldots,N-1$, and $\ell_N=0$.
$\ell_i$ is just the number of boxes in the ith row in the Young
tableau of the representation $\l$. Notice that, defining \be
l_i=\ell_i-\k+\r_i \ee where the Weyl vector is \be
\r_i={N+1\over2}-i \ee we have \be \sum_{i=1}^Nl_i=0~. \ee In the
new basis, the inner product is now \be
(\l,\m)=\sum_{ij}\l_i\m_j(C^{-1})_{ij}=\sum_i l_im_i~. \ee An
$SU(N)$ character is now: \be\label{character}
\chi_\l(\th_1,\ldots,\th_N)={\det\left(e^{il_i\th_j}\right)_{ij}\over\det\left(e^{i\r_i\th_j}\right)_{ij}}
\ee and \be \sum_{i=1}^N\th_i=\sum_{i=1}^Nl_i=0~. \ee

In this basis, the Casimir is given by: \be
C_2(\l)=\sum_{i=1}^Nl_i^2-{1\over12}\,N(N^2-1)~. \ee The Casimir in
the trivial representation is zero. The Casimir in the fundamental
is \be C_2(\tableau{1})=N-{1\over N}~. \ee A commonly used
convention differs from our by a factor of $\half$. The Casimir for
an antisymmetric representation of $N$-ality $k$ in our conventions is
then
\be
C_2(\mbox{A}(k)) ={N+1\over N}\,k(N-k) ~.
\ee

In the main text we used the following representation of the delta
function on a group: \be \d(U)=\sum_\l{\mbox{dim}}(\l)\,\chi_\l(U)
\ee where $U$ is a group element  $SU(N)$. The dimension is given by
\be \dim(\l)=\prod_{1\leq i<j\leq N}{(\ell_i-\ell_j+j-i)\over(j-i)}
\ee and $\ell_i$ is the number of boxes in the $i$th row of the
Young tableau.

It is easy to see that the center of $SU(N)$ acts on the Cartan angles as
$\th_i$ like $\th_i\rightarrow\th_i+{2\pi\over N}$, for $i=1,\ldots,N$.
To see the effect on the character, we use the determinantal expression
\eq{character}. We rewrite the determinant as:
\be
\det(\th_i)=\det(e^{il_i\th_j})_{ij}=\sum_{\s\in
S_N}\e(\s)\prod_{i=1}^Ne^{il_{\s(i)}\th_i} \ee where
$\th_N=-(\th_1+\ldots+\th_{N-1})$. Shifting the determinant by the
constant shifts gives: \bea \det(\th_i+{2\pi\over N})&=&\sum_{\s\in
S_N}\e(\s)e^{{2\pi i\over N}\,l_{\s(1)}+{2\pi i\over N}\,l_{\s(2)}+
\ldots+{2\pi i\over N}\,l_{\s(N-1)} -{2\pi i\over
N}(N-1)l_{\s(N)}}\prod_{i=1}^Ne^{il_{\s(i)}\th_i}\nn &=&\sum_{\s\in
S_N}\e(\s)e^{{2\pi i\over N}\sum_{i=1}^Nl_{\s(i)}}e^{-2\pi
il_{\s(N)}}\prod_{i=1}^Ne^{il_{\s(i)}\th_i} \eea Notice that the
first factor vanishes because $\sum_{i=1}^Nl_i=0$. The second factor
gives: \be e^{-2\pi il_i}=e^{-2\pi i(\k+{N+1\over 2})} \ee for any
$i=1,\ldots,N-1$. Therefore, we get \be \det(\th_i+{2\pi\over
N})=e^{2\pi i k/N}e^{-\pi i(N+1)}\det(\th_i) \ee where $k=N\k$ is the
$N$-ality. We
can now do the same in the denominator by setting $\k=0$. Thus: \be
\det(e^{i\rho_i(\th_j+{2\pi\over N})}=e^{-\pi
i(N+1)}\det(e^{i\rho_i\th_j})~, \ee Therefore the minus signs
cancel, and we are left with \be \chi_\l(\th_i+{2\pi\over
N})=e^{2\pi ik/N}\chi_\l(\th_i) \ee which is what we wanted to prove.

\section{Properties of the interacting theory}\label{interaction}
We can use orthogonality of $SU(N)$ characters of irreducible
representations: \be
\bra\chi_\m,\chi_\n\ket=\int\dd\th\,J^2(\th)\,\chi_{\bar\m}(\th)\chi_\n(\th)=\d_{\m\n}~.
\ee The inner product is defined with respect to the Haar measure,
as usual. We now also require that the $\psi_\l$'s are orthonormal:
\be \bra\psi_\l,\psi_\m\ket=\d_{\l\m}~. \ee The same is true for the
$\o$'s. This implies \be \sum_{\m\atop
k_\m=k_\l=k_{\l'}}c^{\bar\m}_{\bar\l}c^\m_{\l'}=\sum_{\m\atop
k_\m=k_\l=k_{\l'}}\o^{\bar\m}_{\bar\l}\o^\m_{\l'}=\d_{\l\l'} \ee so
that \be c^{\bar\m}_{\bar\l}=\o^\l_\m=(c^{-1})^\l_\m~, \ee so $c$ is
a unitary matrix.

Having introduced $g$ and $n$, a natural question to ask is: if we
are free to choose the $u$'s, how many components of $n$ can we
choose? Notice that \be n^\m_0=u_\m \ee where $0$ is the trivial
representation. Therefore, choosing $u_\m$ is equivalent to choosing
$n^\m_0$. Once this is done, all other components of $n^\m_\n$ are
fixed.

We can use the charge conjugation matrix to lower indices: \be
c_{\m\bar\n}\equiv c_\m^\n~. \ee Indeed, this definition is correct
because 
\be 
N_{\m\n}^{\bar\l}=N_{\m\n\l}=\int[d\th]\,\chi_\m(\th)\chi_\n(\th)\chi_\l(\th)~.
\ee 
It therefore satisfies \be N_{\m\n}^{\bar\l}=N_{\m\l}^{\bar\n}~.
\ee
From this, we derive
\be 
n_{\m\n}=n_{\n\m}=n_\m^{\bar\n} \ee and \be
n_\m^\n=n_{\bar\n}^{\bar\m}~. \ee Notice that the newly defined
eigenfunctions $\psi$ also satisfy: 
\bea
\psi_\m(\th)\psi_\n(\th)&=&\sum_\l \ti N^\l_{\m\n}\psi_\l(\th)\nn
\ti N^\l_{\m\n}&=&\int[\dd\th]\,\psi_\m(\th)\psi_\n(\th)\psi_{\bar\l}(\th)~. 
\eea 
The $\ti N$'s can
be computed in terms of the tensor multiplicity coefficients: 
\be
\ti N^\l_{\m\n}=\sum_{\s\s'} c_\m^\s c_\n^{\s'} N_{\s\s'}^\l~. 
\ee
Finally, we should ensure positive definiteness of the potential.
This means \be \bra\psi_\m|U|\psi_\n\ket=\sum_{\l\l'\s}\o^\m_\l
c^{\l'}_\n n^\l_{\l'}\geq 0~. \ee This ensures that the energy is
positive. Indeed, from the above we get for the additional term in
the energy: \be \d E_\l\geq 0~. 
\ee 


\end{document}